\documentclass[%
 aip,
 amsmath,amssymb,
 reprint,%
]{revtex4-1}

\usepackage{graphicx}
\usepackage{dcolumn}
\usepackage{bm}
\usepackage{mathtools}
\usepackage{siunitx}
\usepackage{upgreek}

\makeatletter
\def\@email#1#2{%
 \endgroup
 \patchcmd{\titleblock@produce}
  {\frontmatter@RRAPformat}
  {\frontmatter@RRAPformat{\produce@RRAP{*#1\href{mailto:#2}{#2}}}\frontmatter@RRAPformat}
  {}{}
}%
\makeatother
\begin{document}

\preprint{AIP/123-QED}

\title{Gradient-Free Inverse Design of 3D Nanophotonic Devices Using the Level-Set-Method}

\author{Oliver Kuster}
 \email{oliver.kuster@kit.edu}
 \affiliation{Institute of Theoretical Solid State Physics, Karlsruhe Institute of Technology Kaiserstraße~12, 76131 Karlsruhe, Germany}

\author{Carsten Rockstuhl}%
\affiliation{Institute of Theoretical Solid State Physics, Karlsruhe Institute of Technology Kaiserstraße~12, 76131 Karlsruhe, Germany}
\affiliation{Institute of Nanotechnology, Karlsruhe Institute of Technology, Kaiserstraße~12, 76131 Karlsruhe, Germany}
\affiliation{Center for Integrated Quantum Science and Technology, Karlsruhe Institute of Technology, Wolfgang-Gaede-Str.~1, 76131 Karlsruhe, Germany}

\author{Thomas Jebb Sturges}%
\affiliation{Institute of Nanotechnology, Karlsruhe Institute of Technology, Kaiserstraße~12, 76131 Karlsruhe, Germany}

\date{\today}

\begin{abstract}
Advances in 3D nanophotonic additive manufacturing enable increasingly complex, almost free-form devices for manipulating light at sub-wavelength scales.
Extending nanophotonic designs into three dimensions introduces additional degrees of freedom.
Exploiting this freedom requires inverse-design methods that can efficiently navigate this large design space while accommodating fabrication constraints.
Here, we combine gradient-free optimization with the level-set method (LSM) to provide such an approach for 3D additive manufacturing.
Representing the 3D geometry as a level set constructed from Gaussian radial basis functions retains substantial geometric freedom with only a few hundred design variables.
This compact representation makes gradient-free optimization practical for fully 3D nanophotonic devices, eliminating the need for adjoint calculations. This unlocks the use of any Maxwell solver or auxiliary algorithms that should express fabrication constraints for which the required derivatives are unavailable or impractical to obtain, and enables embarrassingly parallel workflows.
Consequently, binary fabrication requirements, such as structural connectivity, can be rigorously enforced, avoiding the differentiable approximations commonly used in gradient-based optimization.
Here, we demonstrate the combination of the gradient-free LSM with a connected-component analysis algorithm to enforce the structural integrity of the design.
We demonstrate the applicability of the gradient-free LSM by optimizing a Y-splitter while strictly enforcing the structural integrity of the material and the void in the design.
\end{abstract}

\maketitle

\section{\label{sec:introduction}Introduction}
Free-form 3D additive manufacturing for nanophotonics allows for the printing of almost any conceivable design in all three spatial dimensions \cite{fischerThreedimensionalOpticalLaser2013, somersPhysics3DPrinting2023, hahnRapidAssemblySmall2020, kieferMultiphoton77focus2024}.
By exploiting the third dimension, the additional degrees of freedom enable the encoding of complex optical responses in compact devices at the micrometer length scale and below.
In turn, the design process of free-form 3D nanophotonic devices requires efficient algorithms to find optimized designs, ideally while also considering constraints that are put on the design given by the fabrication method.
Inverse design has emerged as a powerful tool to optimize nanophotonic devices. By encoding the desired functionality into a figure of merit, the material distribution of the nanophotonic device can be optimized, iteratively improving the figure of merit \cite{moleskyInverseDesignNanophotonics2018a, christiansenInverseDesignPhotonics2021a, parkFreeformOptimizationNanophotonic2022}.
The material distribution itself is determined by the parametrization of the problem. Popular choices for the parametrization include shape optimization \cite{schneiderBenchmarkingFiveGlobal2019, liuBreakingSizeToleranceTradeoff2026, liCompactBroadbandMultimode2020, razaFabricationawareInverseDesign2025a, michaelsHierarchicalDesignOptimization2020}, density-based methods \cite{hughesAdjointMethodInverse2018a, huangCompactInverseDesigned2025, wangInverseDesignOptical2026, christiansenOrdersMagnitudeReduction2026a, hammondTopologyoptimizedDistributed3d2026a, shiUltracompactMultimodeParallel2026}, and the level-set method (LSM) \cite{burgerFrameworkConstructionLevel2003, burgerSurveyLevelSet2005, voronovInversedesignTopologyOptimization2025, vercruysseAnalyticalLevelSet2019, piggottFabricationconstrainedNanophotonicInverse2017b, kaoMaximizingBandGaps2005, lebbeRobustShapeTopology2019, saad-falconLevelSetMethods2024, kimAutomatedDiscoveryOptimization2023}.

In nanophotonics, density-based methods have shown great success and established themselves as one of the most common methods used.
Density-based methods assign a continuous material variable to each spatial cell in a discretized computational grid and thereby allow for free-form designs. The resulting problems typically have thousands to millions of degrees of freedom and are commonly optimized using the adjoint method \cite{christiansenInverseDesignPhotonics2021a}. The continuous distribution must ultimately be converted into a binary, fabricable design. Filters and projections can promote binarization during the optimization, but increasingly sharp projections suppress sensitivities throughout much of the design region and can cause the optimization to stall. Simply thresholding the final distribution, on the other hand, can degrade its optical performance if the design relied too much on non-binarized regions.

In contrast, the LSM instead represents the material boundary as a contour of a scalar function, maintaining a binary material distribution with a clearly defined interface. The scalar function is typically represented as a sum of basis functions, allowing complex structures to be described using only a few hundred design variables. The LSM is often combined with adjoint-based optimization, in which the geometry is updated by evolving its boundaries. However, changes in topology can then be difficult to achieve, making the result sensitive to the initial topology and often requiring a good initial guess.

For problems with only tens to a few hundred degrees of freedom, gradient-free and global optimization methods provide a useful alternative to adjoint formulations \cite{jafar-zanjaniAdaptiveGeneticAlgorithm2018, flanneryFabryPerotCavityFormed2018, yuGeneticalgorithmoptimizedWidebandOnchip2017, liInverseDesignMultifunctional2019, qiuOptimizationBroadbandOptical2012, huangOptimizationPhotonicNanojets2019, chungTunableMetasurfaceInverse2020, parkUltimateLightTrapping2019}. The compact basis-function representation of the LSM makes such approaches feasible, allowing the optimizer to explore changes in topology that are difficult to reach through gradual boundary evolution. Avoiding gradients also enables the direct use of non-differentiable fabrication constraints.
This is particularly useful for 3D nanophotonic design, where both the material and, typically, the void must remain connected.
3D laser nanoprinting is a popular fabrication method for printing three-dimensional photonic devices at the micrometer scale. By focusing a laser inside a photoresist, a non-linear absorption process can trigger polymerization precisely at the focal spot of the laser.
By controlling the path the laser takes, it is possible to obtain full 3D nanophotonic devices. The non-polymerized photoresist is then removed during development, leaving only the full 3D nanophotonic device \cite{fischerThreedimensionalOpticalLaser2013, somersPhysics3DPrinting2023}.
However, an additive 3D printing process naturally also requires connectivity of the design and the void regions.
Connectivity of the material guarantees that no disconnected parts appear during development, which would otherwise be washed away during development or simply fall off. Furthermore, void connectivity is also required, as enclosed voids trap the photoresist and can alter the electromagnetic response of the final device.
Connectivity is a binary measure and therefore impossible to rigorously impose with a gradient-based optimization scheme. Instead, a differentiable approximation must be used that only promotes connectivity. This makes it challenging to find a connected structure without either overconstraining the design or accepting some amount of free-floating or disconnected parts \cite{coolPracticalReviewPromoting2025a}.

\begin{figure*}[t]
    \centering
    \includegraphics[width=\linewidth]{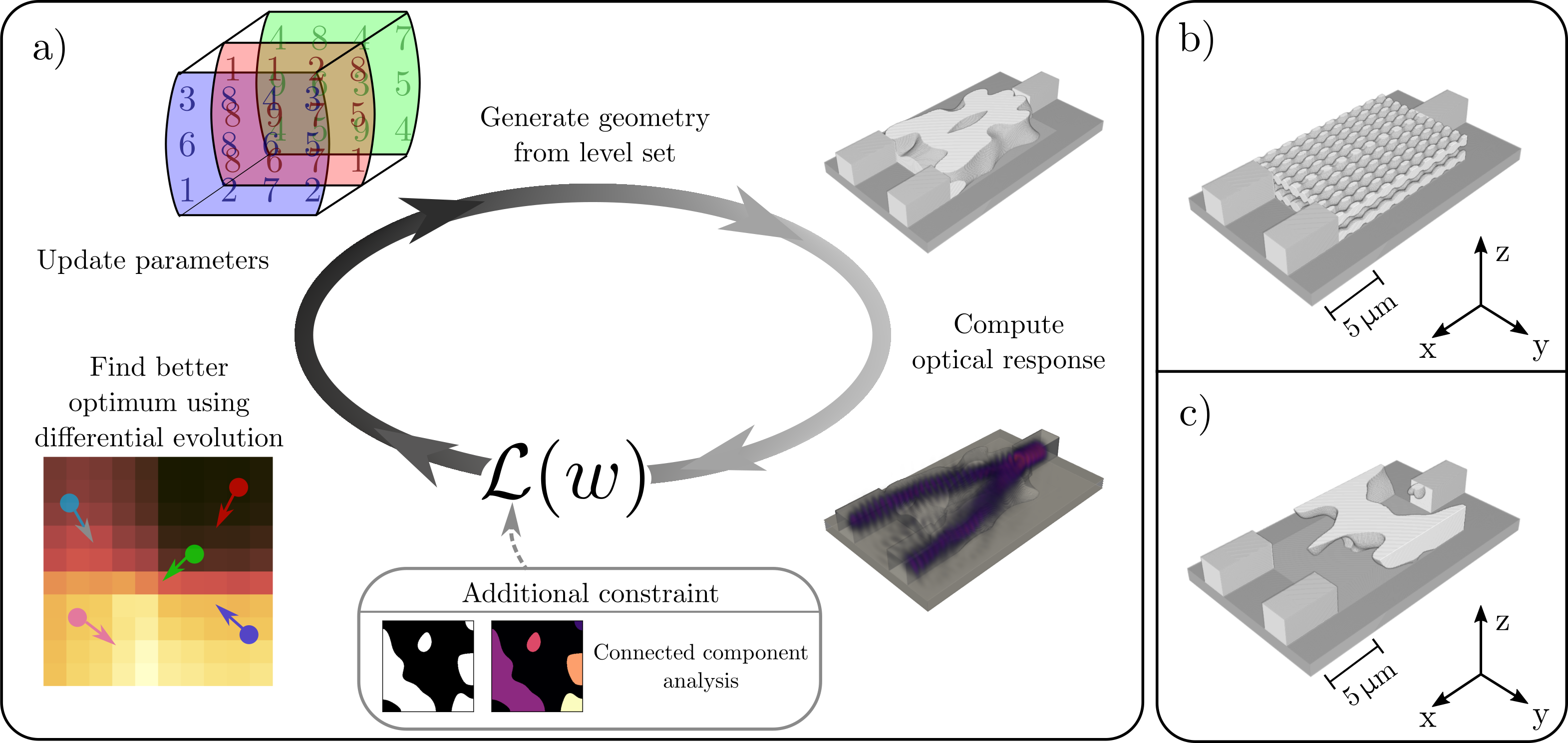}
    \caption{a) Sketch of our gradient-free level set method. (i) The parameterization can be represented as a matrix of amplitudes of the Gaussian radial basis functions (RBFs) that are distributed uniformly throughout the design region. (ii) The device geometry is obtained by taking the level set of these Gaussians. (iii) We use a finite-difference time-domain solver to compute the optical response of the design, which defines the electromagnetic figure of merit $\mathcal{L}_\text{EM}(\mathbf{p})$. (iv) We then include additional constraints on our design, such as structural integrity using connected component analysis, to calculate the full figure of merit $\mathcal{L}(\mathbf{p})$. (v) By using a differential evolution algorithm, we iteratively update the weight matrix to gradually optimize our design. b) An example design with uniformly initialized amplitudes of the Gaussian RBFs. c) An example design with randomly initialized amplitudes of the Gaussian RBFs. Both designs in b) and c) use $12$ Gaussians in the $x$-direction, $8$ Gaussians in the $y$-direction, and $4$ Gaussians in the $z$-direction, resulting in $N=12\times 8\times 4 = 384$ weights, and use $\sigma=\SI{0.5}{\upmu m}$ for all Gaussian radial basis functions.}
    \label{fig:cover_pic}
\end{figure*}

In this work, we combine gradient-free optimization with an LSM parametrization and apply the method to 3D nanophotonic design. Additionally, we guarantee material and void connectivity of the 3D nanophotonic device. The gradient-free LSM has previously been demonstrated for 2D metasurfaces \cite{saad-falconLevelSetMethods2024} and 3D photonic crystals \cite{kimAutomatedDiscoveryOptimization2023}, but did not consider any rigorous constraints in their optimization. Here, we explore a free-form design in all three spatial dimensions and enforce connectivity as a strict constraint to make it suitable for 3D additive manufacturing. Independent candidate designs can be evaluated in parallel, and any suitable Maxwell solver can be used without implementing the adjoint formalism.
A sketch of the optimization workflow can be found in Fig.~\ref{fig:cover_pic} a).

\section{\label{sec:methods}Methods}
The general inverse design problem in nanophotonics using a time-domain solver can be formulated as
\begin{align}
  &\max\,\,\,\mathcal{L}(\mathbf{D}(\mathbf{r}, t;\mathbf{p}) =: \mathcal{L}(\mathbf{p}), 
  \label{eq:to_problem}\\
  &\text{s.t.}  \,\,\,  \frac{\partial\mathbf{B}(\mathbf{r}, t)}{\partial t} = - \nabla \times \mathbf{E}(\mathbf{r}, t) \nonumber\\
  &\text{s.t.}  \,\,\, \frac{\partial\mathbf{D}(\mathbf{r}, t; \mathbf{p})}{\partial t} = \nabla \times \mathbf{H}(\mathbf{r}, t) - \mathbf{J}(\mathbf{r}, t),\nonumber\\
  &\text{s.t.}  \,\,\, \mathbf{c} = \mathbf{0},\nonumber\\
  &\text{s.t.}  \,\,\, \mathbf{d} < \mathbf{0},\nonumber
\end{align}
where $\mathbf{p}$ is the parametrization of the problem, and $\mathcal{L}(\mathbf{p})$ the figure of merit. The time evolution of Maxwell's equations enters as an equality constraint. With $\mathbf{E}(\mathbf{r},t)$ representing the electric field and since we assume non-dispersive material, we can write the electric displacement field as $\mathbf{D}(\mathbf{r}, t; \mathbf{p}) = \varepsilon_0\varepsilon(\mathbf{r}, \mathbf{p})\mathbf{E}(\mathbf{r}. t)$. $\mathbf{D}(\mathbf{r}, t; \mathbf{p})$ encodes the optical response to the material distribution resulting from the parametrization $\mathbf{p}$. The magnetic flux density are given as $\mathbf{B}(\mathbf{r}, t)$, and the magnetic field can be written as $\mathbf{H}(\mathbf{r}, t) = \frac{1}{\mu_0}\mathbf{B}(\mathbf{r}, t)$. We assume that our material has no magnetic response and omit the parametrization in our magnetic flux density. The source is given as $\mathbf{J}(\mathbf{r},t)$.

Design requirements can be implemented by restricting the feasible design space through equality constraints $\mathbf{c} = \mathbf{0}$ or inequality constraints $\mathbf{d} < \mathbf{0}$, or through penalty or regularization terms in the figure of merit $\mathcal{L}(\mathbf{p})$. 
Regardless of the formulation, gradient-based optimization requires differentiable constraints.
This can make discrete fabrication requirements difficult to enforce directly, often requiring differentiable approximations instead.

A particularly important requirement for full 3D designs is structural integrity, for which structural connectivity is a fundamental condition. Connectivity is a binary measure. Either structures are connected or not connected, making connectivity a non-differentiable function.
Implementing connectivity into a differentiable framework has led to substantial research into differentiable approximations to structural connectivity~\cite{coolPracticalReviewPromoting2025a}.
However, constraining such an approximation does not in general guarantee true connectivity.
Moreover, when such approximations are incorporated as penalty terms, their weighting introduces a trade-off between satisfying the fabrication requirement and optimizing the photonic performance.
Gradient-free optimization, by contrast, can evaluate the connectivity condition directly, irrespective of its differentiability.

Within our gradient-free LSM approach, we use connected-component analysis to reject disconnected structures and thereby rigorously enforce structural integrity throughout the optimization.
Connected-component analysis works on arrays with two distinct entries (e.g., a binarized image), where entries are separated into foreground and background pixels. 
The array is converted into a linked list, where successively neighboring foreground pixels are given the same label. Thus, a fully connected structure will only return a single label, which we use to distinguish between a fully connected and a non-fully connected structure.
The gradient-free formulation of our problem makes the use of a connected-component analysis a very easy and convenient algorithm to implement for connectivity constraints.

Using connected-component analysis, we verify that both the material and void regions each form a single connected region, ensuring a fully connected structure without isolated material features or enclosed cavities. A design is considered feasible only if both connectivity requirements are satisfied. The final figure of merit is defined using a feasibility-based constraint-handling strategy \cite{coellocoelloTheoreticalNumericalConstrainthandling2002} as
\begin{equation}
\mathcal{L}(\mathbf{p}) =
\begin{cases}
\mathcal{L}_{\mathrm{EM}}(\mathbf{p}),
& \text{if feasible},\\
0,
& \text{otherwise}.
\end{cases}
\end{equation}
The optical performance of the device is evaluated using $\mathcal{L}_\text{EM}(\mathbf{p)}$, and feasibility is measured by whether or not the material or void returns more than one label in the connected-component analysis.
In this formulation, connectivity is treated as a strict feasibility constraint: any design violating either requirement is assigned a figure of merit of zero, whereas $\mathcal{L}_{\mathrm{EM}}(\mathbf{p})$ is always strictly non-negative.

In the LSM, the geometry is represented implicitly by a scalar field $f(\mathbf{r};\mathbf{p})$ whose level set defines the boundary between material and void.
For a chosen threshold $\Delta$, this boundary is given by $f(\mathbf{r};\mathbf{p}) = \Delta$.
The scalar field is parameterized as a weighted sum of basis functions $f(\mathbf{r};\mathbf{p}) = \sum_{i=0}^N p_i \phi_i(\mathbf{r})$. 
Here, we choose the Gaussian radial basis functions (RBFs) $\phi_i(\mathbf{r}) = \exp\left(-\frac{(\mathbf{r}-\boldsymbol{\mu}_i)^2}{\sigma_i^2}\right)$, where $\boldsymbol{\mu}_i$ represents the centers of each Gaussian, and $\sigma_i$ represents the standard deviation of each Gaussian. The feature sizes obtained by the LSM mostly depend on the total number of weights and the values of $\sigma_i$.
For simplicity, we distribute the Gaussians uniformly on the grid of the design region and leave $\boldsymbol{\mu}_i$ and $\sigma_i =: \sigma$ constant during the optimization.
This means that the entire parametrization for our optimization is encoded in the amplitudes of the Gaussians $p_i$.
Figure~\ref{fig:cover_pic} b) shows the design with a uniform parameterization $p_i = p_0$ for all $i\in [0, N]$, visually demonstrating how the level set of the 3D Gaussian RBFs gives rise to overlapping spheroids.
Figure~\ref{fig:cover_pic} c) shows randomly chosen $p_i$, providing an example of the level of geometric complexity that can be achieved with our parameterization.
The designs were parametrized using $12$ Gaussians in the $x$-direction, $8$ Gaussians in the $y$-direction, and $4$ Gaussians in the $z$-direction, resulting in $N=384$ weights.

For optimization, we use an evolutionary algorithm, as it is well suited to problems with several hundred degrees of freedom.
Evolutionary algorithms randomly generate a randomized initial population $\mathbf{p}_k$ $k\in [1, M]$ with a size $M$ as their first generation. By evaluating $\mathcal{L}(\mathbf{p}_k)$ for each individual $k$ in the population, the best-performing individuals of that generation can be found. 
The next generation is then generated by updating and recombining the population parameters based on the best-performing individuals.
Iteratively, each new generation will tend to perform better than the last one, successively improving $\mathcal{L}(\mathbf{p})$ over the optimization procedure.
Our choice of evolutionary algorithm is a differential evolution algorithm \cite{connIntroductionDerivativefreeOptimization2009, stornDifferentialEvolutionSimple1997}.

Lastly, since the design is parameterized by the weights of continuous Gaussian RBFs, its geometry is defined independently of the discretization used by the electromagnetic solver. The same optimized parameters can therefore be evaluated directly on a finer grid, without interpolating or otherwise modifying the design. This reduces the optimization cost by using a coarser grid, while the final design can then be verified at higher resolution using the same geometry and parameterization.
Nevertheless, for the performance obtained during optimization to carry over to the finer-grid evaluation, the optimization grid must still be sufficiently fine to capture the relevant electromagnetic response. If the optimization grid is too coarse, the optimizer may exploit discretization errors, resulting in a design whose performance deteriorates when evaluated at a higher resolution.

\section{\label{sec:results}Optimization of a Y-Splitter}
We demonstrate the gradient-free LSM optimization on a 3D printable Y-splitter sitting on top of a glass substrate. 
The design goal is to split the light from an input waveguide evenly between two output waveguides.
To demonstrate a fully three-dimensional design, we consider a structure suitable for fabrication by 3D laser nanoprinting.
We model the printed device as a polymer resin with a nondispersive relative permittivity $\varepsilon_\text{material}=1.55^2$, supported by a glass substrate with $\varepsilon_\text{sub}=1.44^2$, and surrounded by air with $\varepsilon_\text{air}=1$.
Furthermore, we choose $\lambda_0 = \SI{1.55}{\upmu m}$ as the operational wavelength, resulting in $\lambda_\text{material}=\SI{1}{\upmu m}$ as the effective wavelength inside the polymer resin.
We choose rectangular waveguides with a relatively large cross section of $\SI{2}{\upmu m}\times\SI{2}{\upmu m}$ to ensure sufficient confinement despite the low refractive-index contrast between the waveguide material and the substrate.
The design region is matched in height to the waveguides and has a total size of $\SI{12}{\upmu m}\times\SI{8}{\upmu m}\times\SI{2}{\upmu m}$.
Furthermore, we use mirror symmetry along the $y$-axis to reduce the computational domain for the simulation.
The simulations are performed using the finite-difference time-domain Maxwell solver Meep \cite{oskooiMeepFlexibleFreesoftware2010} and discretized on a grid using $\SI{10}{\text{vx}/\upmu m}$.

We parametrize the optimization by choosing a weight matrix $\mathbf{p}$ of size $12\times 8 \times 4$, resulting in roughly one Gaussian per micrometer in the $x$-direction and two Gaussians per micrometer in the $y$- and $z$-direction after symmetry considerations. The total number of adjustable weights is thus $384$.
We use a uniform $\sigma=\SI{1.5}{\upmu m}$ for all Gaussians in our RBF.
We choose $\sigma$ to be on the larger side, comparable to the wavelength, since we want to promote designs whose features are not too small w.r.t the wavelength, making them easier to fabricate.

We excite the fundamental mode of the input waveguide — which is a transverse magnetic mode — and maximize its coupling efficiency into the fundamental mode (TM$_{00}$) of the output waveguide, defining our electromagnetic figure of merit $\mathcal{L}_\text{EM}(\mathbf{p})$.
As mentioned earlier, connectivity of material and void are strictly enforced during the optimization. A gradient-free formulation makes the optimization particularly easy, as it requires no differentiability or continuity of the figure of merit when using evolutionary algorithms.

\begin{figure}
    \centering
    \includegraphics[width=\linewidth]{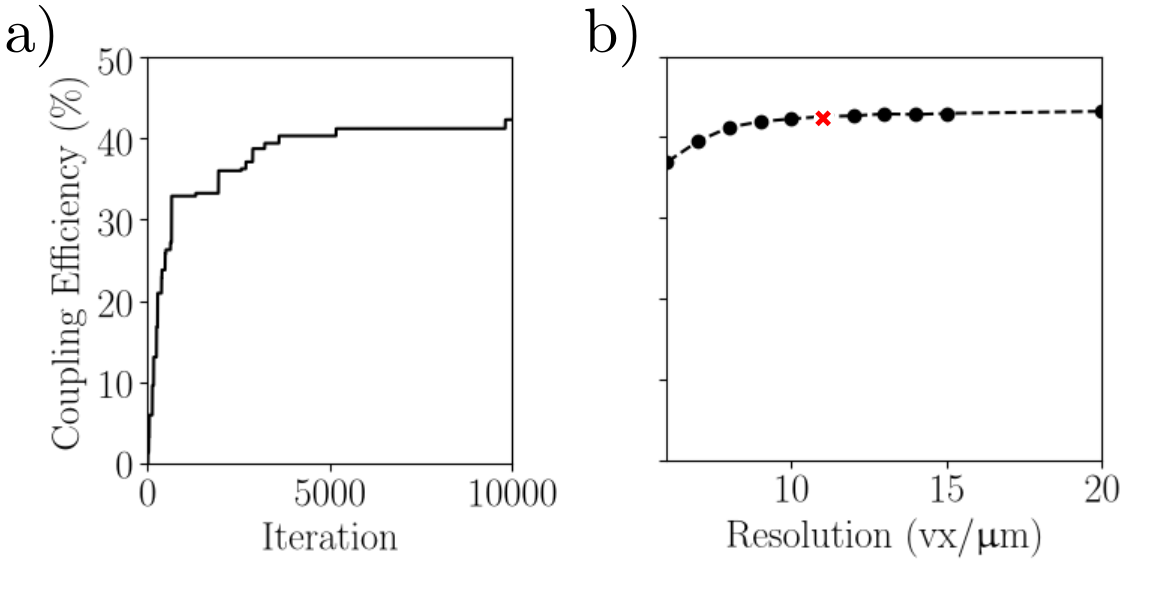}
    \caption{a) The optimization history of the coupling efficiency of the fundamental mode into one output waveguide of the Y-splitter. The optimization ran for $10\,000$ iterations and reached a coupling efficiency of $42.5\%$.
    b) The dependence of the figure of merit on the resolution of an already optimized Y-splitter. The Y-splitter was optimized at a resolution of $\SI{10}{\text{vx}/\upmu m}$ as indicated by the red cross. We then use the optimized weight matrix $\mathbf{p}$ and evaluate the figure of merit of the optimized Y-splitter at different resolutions.}
    \label{fig:loss_plot}
\end{figure}

\begin{figure*}[t]
    \centering
    \includegraphics[width=\linewidth]{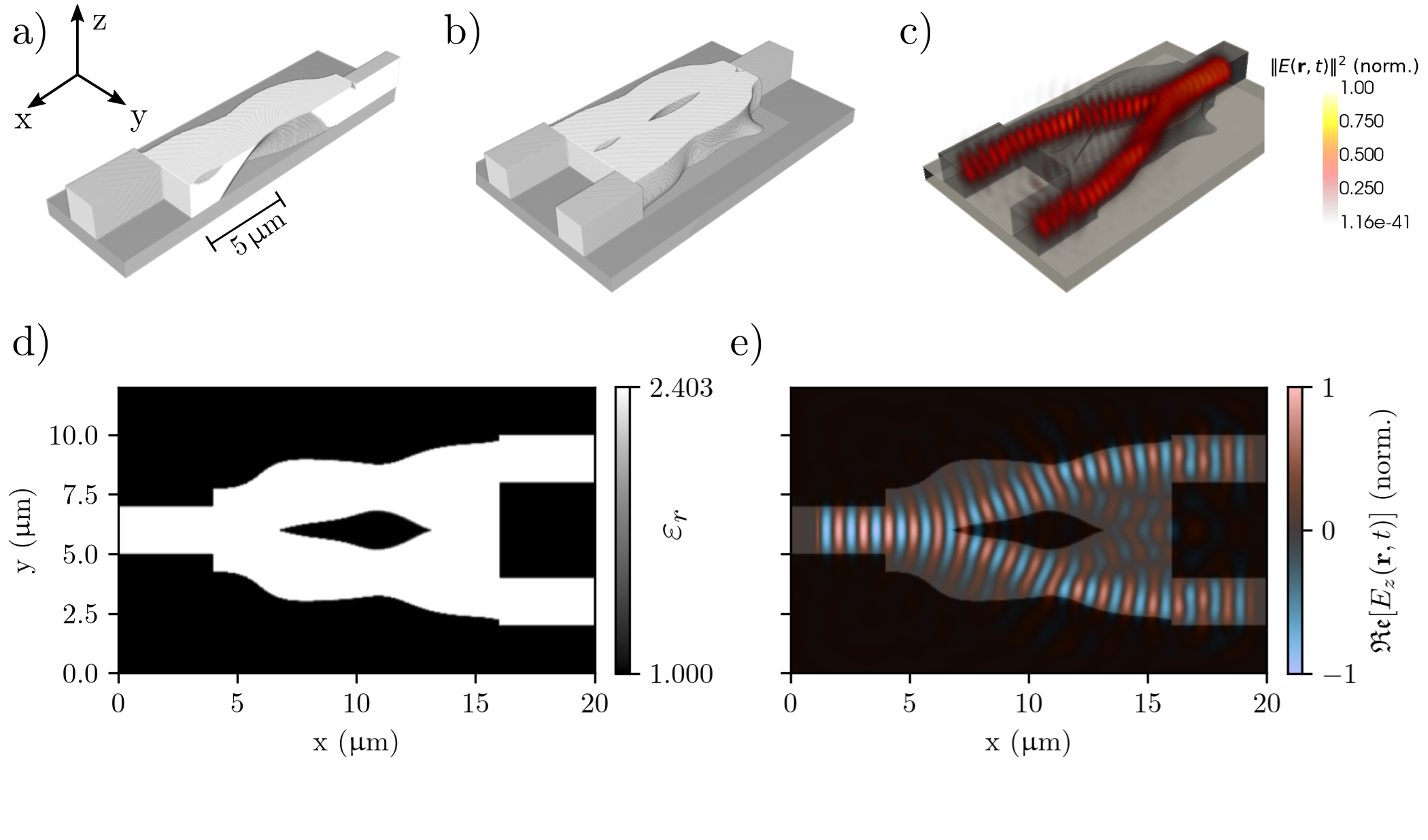}
    \caption{The final, best-performing Y-splitter and its optical performance.
    a) Shows the permittivity distribution of half of the Y-splitter and its interior geometry. Due to symmetry, we simulate only half of the Y-splitter in our optimization. b) Shows the full Y-splitter, using symmetry along its y-axis, with its input and output waveguides on top of a substrate. The Y-splitter is assumed to have a relative permittivity of $\varepsilon_\text{material} = 1.55^2$ to represent the relative permittivity of typical polymers used in 3D laser-nanoprinting. The substrate is assumed to be made of glass $\varepsilon_\text{sub}=1.44^2$. The design is optimized using the gradient-free level-set method presented in this work. Additionally, the structure is fully connected in material and void, which we added as a rigorous constraint for our optimization. c) Shows the intensity distribution of the Y-splitter when excited through the input port.
    d) Shows the 2D cross section along the z-axis at the middle of the waveguide of the Y-splitter, and e) the real part of the $z$-polarized electric field $\mathfrak{Re}[E_z(\mathbf{r}, t)]$ in the same cross section as shown in d).
    }
    \label{fig:field_and_structure}
\end{figure*}

We run the optimization for $10\,000$ iterations, using Meta's Nevergrad optimization library \cite{jeremyNevergradGradientfreeOptimization2018}, which automatically chooses and adapts the hyperparameters (population size, number of parents, etc.) of the gradient-free optimizer. In particular, we use its ``TwoPointsDE'' as our optimization algorithm.
The full evolution of the coupling efficiency over the optimization can be seen in Fig.~\ref{fig:loss_plot} a).
The final, optimized design can be seen in Fig.~\ref{fig:field_and_structure} a) and b).  The intensity distribution and optical response of the final, optimized design can be seen in Fig.~\ref{fig:field_and_structure} c). Furthermore, we also show a 2D cross section of our design and the real part of the electric field in Fig.~\ref{fig:field_and_structure} d) and e), respectively.
The design achieves a coupling efficiency of $42.5\%$ for the waveguide's fundamental mode and a total flux transmission of more than $45\%$ into the output waveguide. We note that the coupling efficiency and total flux transmission are given for a single output waveguide.

We also note that the performance of the Y-splitter is evaluated on a finer computational grid ($\SI{20}{vx/\upmu m}$) than that used during optimization ($\SI{10}{vx/\upmu m}$). 
To verify that resolution does not significantly affect device performance, we calculate the coupling efficiency of the design obtained on a grid with a resolution of $\SI{10}{\text{vx}/\upmu m}$ at different resolutions. The resulting coupling efficiencies can be seen in Fig.~\ref{fig:loss_plot} b), where we can see that the final design (represented by a red cross) performs just as well at higher resolutions.

The optimization was performed with $8$ processes in parallel on an AMD EPYC 7413 CPU, resulting in a total optimization time of roughly $\SI{75}{h}$ for the $10\,000$ iterations.

\section{\label{sec:conclusion}Conclusion}
We present a gradient-free optimization approach to free-form 3D nanophotonic design based on the LSM.
Using connected-component analysis, we enforce strict material connectivity and the absence of enclosed cavities, making the resulting structures well suited to fabrication by 3D laser nanoprinting.
We demonstrate the approach by designing a Y-splitter with a footprint of $\SI{12}{\upmu m}\times\SI{8}{\upmu m}\times\SI{2}{\upmu m}$, which can reach a coupling efficiency of $42.5\%$ into the fundamental mode and a flux transmission of more than $45\%$ into the output waveguide.
The relatively low dimensionality of the LSM parameterization enables the use of a differential evolution algorithm despite the fully three-dimensional design space.
Compared with previous applications of gradient-free LSM optimization, our approach considers a fully three-dimensional, effectively free-form geometry and explicitly incorporates connectivity requirements as strict feasibility constraints.
The optimization is implemented in a fully open-source manner and is easily reproducible, using Nevergrad for the optimization algorithm, Meep as our Maxwell solver, and SciPy for connected-component analysis.

We acknowledge that gradient-based optimization remains considerably more computationally efficient for many large-scale nanophotonic inverse-design problems and can therefore operate in much higher-dimensional design spaces. However, the gradient-free LSM considered here offers complementary advantages: it requires neither a differentiable electromagnetic solver nor differentiable fabrication constraints, permits discrete geometric requirements such as connectivity to be enforced directly, and naturally restricts the optimization to a comparatively low-dimensional family of smooth geometries. For devices such as the Y-splitter considered here, the gradient-free LSM can favor comparatively simple structures that guide the optical field without relying strongly on intricate sub-wavelength features.

Our results demonstrate that gradient-free LSM optimization provides a simple and flexible route for free-form 3D nanophotonic inverse design, particularly in settings where fabrication requirements are difficult to express through differentiable constraints.

\section{Acknowledgments}
O. K. and C. R. acknowledge support through the Deutsche Forschungsgemeinschaft (DFG, German Research Foundation) under Germany’s Excellence Strategy via the Excellence Cluster 3D Matter Made to Order (EXC-2082/2, Grant No. 390761711) and from the Carl Zeiss Foundation via CZF-Focus@HEiKA. C. R. and T. J. S. acknowledge support by the Helmholtz Association via the Helmholtz program “Materials Systems Engineering” (MSE). 

\section{Data Availability Statement}
Data underlying the results presented in this paper are not publicly available but may be obtained from the authors upon reasonable request or generated from the publicly available code under \url{https://github.com/OlloKuster/gradient_free_LSM}.

\bibliography{article}
\end{document}